\documentclass[onecolumn,amsmath,amssymb,subfigure,12pt]{revtex4-1}
\usepackage{bm}
\usepackage{graphicx}
\usepackage{hhline}
\usepackage[newitem,newenum]{paralist}
 {\pointedenum\begin{enumerate}}%
 {\end{enumerate}}
 {\pointlessenum\begin{enumerate}}%
 {\end{enumerate}} 
\usepackage{mathrsfs}
\usepackage{amssymb}
\usepackage{bm}
\usepackage[dvips]{color}
\usepackage{color}
\usepackage{fancybox}
\usepackage{fancybox}
\usepackage{fancyhdr}
\begin{document}

\title{1D inelastic relativistic-collision}

\author{Ryosuke Yano }
\affiliation{Department of Advanced Energy, The University of Tokyo, 5-1-5 Kashiwanoha, Kashiwa, Chiba 277-8561, Japan, Telephone number: +81-4-7136-3858}
\email{yano@k.u-tokyo.ac.jp}
\author{Arnaud Martin}
\affiliation{Sino-French Institute for Nuclear Engineering and Technology, Sun Yat-sen University, Tangjiawan, Xiangzhou, City of Zhuhai 519082, Guangdong province, People's Republic of China}
\email{amphyschim@gmail.com}
\begin{abstract}
In this paper, we discuss one dimensional inelastic relativistic-collisions in the framework of the relativistic kinetic theory. In particular, we focus on the relativistic effects on time evolutions of the temperature and flow velocity under the spatially homogeneous state. As a result of the relativistic effects, we obtain the characteristic time evolutions of the temperature and flow velocity, which are never obtained under the nonrelativistic limit.
\end{abstract}
\maketitle

\section{Introduction}
The relativistic binary collision of two heavy ions in the Relativistic Heavy Ion Collider (RHIC) \cite{RHIC} or Large Hadron Collider (LHC) \cite{LHC} has been significant for understanding of dynamics of quark gluon plasma (QGP) \cite{Hatsuda}. In particular, the kinetic study of QGP has been done to describe the dynamics of QGP in the framework of the relativistic Boltzmann equation \cite{Denicol}, whereas the perfect fluidity is explained in framework of the relativistic hydrodynamics by Hirano and Gyulassy \cite{Hirano}. In a similar way, inelastic relativistic-collisions are described by the inelastic relativistic-Boltzmann equation. Actually, the inelastic relativistic-collision between two partons is significant for understanding of QGP \cite{Wong}. Meanwhile, the characteristics of the inelastic relativistic-Boltzmann equation have not been discussed. In this paper, we consider the characteristics of one dimensional (1D) inelastic relativistic-collisions using 1D inelastic relativistic-Boltzmann equation. We remind that 1D elastic collisions never yield changes of states of partons, because the binary elastic collision means the exchange of the velocity of colliding partons. We, however, consider that 1D inelastic relativistic-collisions are significant as the first step to understanding of inelastic relativistic-collisions, as 1D inelastic collisions have been discussed for the granular gas \cite{Ben-Naim}. The primary aim of this paper is an investigation of relativistic effects on time evolutions of the temperature and flow velocity under the spatially homogeneous state, when the collision is described by the inelastic relativistic-collision. Here, readers must notice that the spatially homogeneous state corresponds to the spatially uniform state of partons. Thus, the spatially homogeneous state allows the non-zero flow velocity. The main reason why we consider 1D inelastic relativistic-collision is that analytical result on the time evolution of the temperature can be obtained using 1D inelastic relativistic-Boltzmann equation, whereas the calculation of the time evolution of the temperature via 3D inelastic relativistic-Boltzmann equation involves mathematical difficulties. The analytical result of the time evolution of the temperature is compared with the numerical result, which is obtained by solving 1D inelastic relativistic-Boltzmann equation on the basis of the direct simulation Monte Carlo (DSMC) method \cite{Bird} \cite{Yano}. Throughout this paper, we assume that partons are composed of hard spheres with equal mass and diameter. Meanwhile, we remind that the collisional differential cross section depends on the momentum transferred between two colliding quarks, and the collisional deflection-angle also depends on the momentum, which is transferred between two colliding partons \cite{Peskin} \cite{Hatsuda}, in the energy-regime of the asymptotic freedom. Additionally, we consider the reaction (inelastic relativistic-collision), $qq \rightarrow qq+g$ ($q$: quark, $g$: gluon), under the thermally relativistic state \cite{Yano}, in which we can assume the asymptotic of freedom, when the thermal energy of quarks is larger than $\Lambda_{\mbox{\tiny{QCD}}}$ \cite{Peskin}, namely, thermally relativistic. We, however, neglect effects of the spin of quarks to simplify our discussions. Finally, Grad's 14 moments \cite{Denicol} are calculated using Eckart's decomposition \cite{Eckart}.
\section{Time evolution of temperature under equilibrium state}
Firstly, we assume that the inelastic relativistic-collision $qq \rightarrow qq+g$ does not have its reverse reaction, namely, $qq+g \rightarrow qq$, whereas we can assume another path to the energy loss by the radiation in $qq$ collisions \cite{Aichelin}. Here, the successive inelastic relativistic-collisions, $qq \rightarrow qq+g$, occur, exclusively. In other words, the reactive collision cross section of $qq+g \rightarrow qq$ is assumed to be zero. Finally, such successive inelastic relativistic-collisions are expressed with the inelastic relativistic-Boltzmann equation.\\  
1D inelastic relativistic-Boltzmann equation is formulated under the spatially homogeneous state as
\begin{eqnarray}
&&p^0\frac{\partial f\left(t,p\right)}{\partial x^0} \nonumber \\
&&=A\int_{-\infty}^{\infty} \left[\frac{1}{\mathcal{J}}f\left(t,p^{\prime\prime}\right) f\left(t,p_\ast^{\prime\prime}\right)-f\left(t,p\right) f\left(t,p_\ast\right)\right] F \frac{dp_\ast}{p^0_\ast}
\end{eqnarray}
where $f\left(t,p\right)$ is the distribution function, in which $t$ is the time, namely, $x^0=t$, $p^0=mc\gamma(\tilde{v})$ ($\gamma(\tilde{v})=1/\sqrt{1-\tilde{v}^2}$: Lorentz factor, $v$: velocity of a quark, $\tilde{v}=v/c$, c: speed of light, m: mass of a quark), and $p=p^1=mv\gamma(\tilde{v})$ ($p^\alpha$: two momentum ($\alpha=0,1$)). In Eq. (1), $\mathcal{J}$ is the Jacobian owing to the inelastic relativistic-collision. The right hand side of Eq. (1) corresponds to inelastic relativistic-collisions, where $F=g_{\o}/\left(p^0p^0_\ast\right)$ ($g_{\o}$: M\o ller's relative velocity) \cite{Cercignani}. $A$ in Eq. (1) defines the total collision cross section. In later discussion, we set $m=1$ and $c=1$ to simplify our discussion (i.e., $v=\tilde{v}$).\\
As a result of the direct inelastic relativistic-collision, momentums of two colliding quarks, namely, $p$ and $p_\ast$, change to $p^\prime$ and $p_\ast^\prime$, which are defined by
\begin{eqnarray}
&&p^\prime=p+\frac{1+\Lambda}{2}\left(p_\ast-p\right)\\ 
&&p_\ast^\prime=p_\ast-\frac{1+\Lambda}{2}\left(p_\ast-p\right),
\end{eqnarray}
where $\Lambda$ is the inelasticity coefficient ($0 \le \Lambda \le 1$). On the other hand, momentums of two colliding quarks, namely, $p^{\prime\prime}$ and $p_\ast^{\prime\prime}$, change to $p$ and $p_\ast$, in which $p^{\prime\prime}$ and $p_\ast^{\prime\prime}$ are defined by
\begin{eqnarray}
&&p^{\prime\prime}=p+\frac{1+\Lambda}{2\Lambda}\left(p_\ast-p\right), \nonumber\\ 
&&p_\ast^{\prime\prime}=p_\ast-\frac{1+\Lambda}{2\Lambda}\left(p_\ast-p\right).
\end{eqnarray}
Consequently, the total momentum is conserved by the inelastic relativistic-collision, whereas the total energy ($E+E_\ast=\sqrt{1+p^2}+\sqrt{1+p_\ast^2}$) is not conserved by the inelastic relativistic-collision. Finally, $\mathcal{J}$ in Eq. (1) is the Jacobian, which is defined by
\begin{eqnarray}
J\equiv |\mbox{det} \partial (p^{\prime\prime},p_\ast^{\prime\prime})/\partial \left(p,p_\ast\right)|^{-1}=\Lambda.
\end{eqnarray}
We fix $\Lambda=0$ in later numerical analysis.\\
The time evolution of $N^\alpha(=\int_{-\infty}^\infty p^\alpha f dp/p^0$) is obtained by multiplying $p^\alpha /p^0$ by both sides of Eq. (1) and integrating over the momentum space $dp/p^0$ as
\begin{eqnarray}
d_t N^{\alpha}&=&\frac{A}{2}\int_{-\infty}^{\infty}\int_{-\infty}^{\infty} \left(\frac{{p^\alpha}^\prime}{{p^0}^\prime}+\frac{{p^\alpha_\ast}^\prime}{{p^0_\ast}^\prime}-\frac{{p^\alpha}}{p^0}-\frac{{p^\alpha_\ast}}{p^0_\ast}\right) f\left(t,p\right) f\left(t,p_\ast\right) F \frac{dp_\ast}{p^0_\ast} \frac{d p}{p^0}.
\end{eqnarray}
In the right hand side of Eq. (6), $\int_{-\infty}^{\infty}\int_{-\infty}^{\infty} \left({p^\alpha}^\prime/{p^0}^\prime+{p^\alpha_\ast}^\prime/{p^0_\ast}^\prime-p^\alpha/p^0-p^\alpha_\ast/p^0_\ast \right) f\left(t,p\right) f\left(t,p_\ast\right) F \frac{dp_\ast}{p^0_\ast} \frac{dp}{p^0}=0$, when $\alpha=0$. Here, we must remind that the number density ($n$) temporally changes in accordance with the change of the flow velocity ($u$) owing to $d_t N^0=d_t(n U^0)=0$, that is defined by two velocity, $U^\alpha=\gamma\left(u\right)\left(1,u\right)$. On the other hand, we obtain $d_t N^1 \neq 0$, because we obtain ${p^1}^\prime/{p^0}^\prime+{p^1_\ast}^\prime/{p^0_\ast}^\prime-p^1/p^0-p^1_\ast/p^0_\ast=u^\prime+u_\ast^\prime-u-u_\ast \neq 0$ from Eqs. (2) and (3).\\ 
The time evolution of the energy-momentum tensor, $T^{\alpha\beta}$ ($=\int_{-\infty}^\infty  p^\alpha p^\beta f dp/p^0$), is obtained by multiplying $p^\alpha p^\beta/p^0$ by both sides of Eq. (1) and integrating over the momentum space $dp/p^0$ as
\begin{eqnarray}
d_t T^{\alpha\beta}&=&\frac{A}{2}\int_{-\infty}^{\infty}\int_{-\infty}^{\infty} \left(\frac{{p^\alpha}^\prime{p^\beta}^\prime}{{p^0}^\prime}+\frac{{p^\alpha_\ast}^\prime {p^\beta_\ast}^\prime}{{p^0_\ast}^\prime}-\frac{{p^\alpha}{p^\beta}}{p^0}-\frac{{p^\alpha_\ast}{p^\beta_\ast}}{p^0_\ast}\right) f\left(t,p\right) f\left(t,p_\ast\right) F \frac{d p_\ast}{p^0_\ast} \frac{d p}{p^0}.
\end{eqnarray}
In the right hand side of Eq. (7),\\
$(A/2)\int_{-\infty}^{\infty}\int_{-\infty}^{\infty} \left({p^\alpha}^\prime{p^\beta}^\prime/{p^0}^\prime+{p^\alpha_\ast}^\prime{p^\beta_\ast}^\prime/{p^0_\ast}^\prime-p^\alpha p^\beta/p^0-p^\alpha_\ast p^\beta_\ast/p^0_\ast \right) f\left(t,p\right) f\left(t,p_\ast\right) F \frac{dp_\ast}{p^0_\ast} \frac{dp}{p^0}$ is equal to 0 from Eqs. (2) and (3), when $\alpha=0$ and $\beta=1$.\\
Consequently, the time evolution of $T^{01}$ is obtained using Eq. (29), when $f=f_{MJ}=n/(2K_1(\chi))\exp(-\chi p^\alpha U_\alpha)$ ($\chi=k\theta/mc^2$: thermally relativistic measure, $k$: Boltzmann constant, $\theta$: temperature, $K_n$: $n$-th order modified Bessel function of the second kind, $f_{MJ}$: one dimensional Maxwell-J$\ddot{\mbox{u}}$ttner function \cite{Kremer}), as
\begin{eqnarray}
d_t T_E^{01}&=&d_t \left(n \frac{K_2(\chi)}{K_1(\chi)}U^0 U^1\right)\nonumber \\
&=&n U^0 d_t \left(U^1 \frac{K_2(\chi)}{K_1(\chi)}\right)=0,
\end{eqnarray}
where we used the relation $d_t N^0=d_t \left(n U^0\right)=0$.\\
From Eq. (8), we obtain
\begin{eqnarray}
U^1(t)=\mathcal{C} \frac{K_1(\chi(t))}{K_2(\chi(t))},
\end{eqnarray}
where $\mathcal{C}=U^1(0) \frac{K_2(\chi(0))}{K_1(\chi(0))}$.\\
$K_1(\chi(t))/K_2(\chi(t))$ decreases, as $\chi$ decreases. As a result, $|u|$ increases, as $\chi$ ($\theta$) increases (decreases), whereas $|u|$ decreases, as $\chi$ $(\theta)$ decreases (increases). Similarly, $\chi$ ($\theta$) increases (decreases), as $|u|$ increases, whereas $\chi$ $(\theta)$ decreases (increases), as $\left|u\right|$ decreases. Of course, $u(t)=0$, when $u(0)=0$. \\\\
\textbf{Remark 2.1}\\\\
\textit{The flow velocity $u$ never be conserved by inelastic relativistic-collisions, even when the total momentum of binary colliding quarks is conserved. Provided that $f=f_{MJ}$, $|u|$ is inversely proportion to the temperature and its vice versa. $\lim_{t \rightarrow \infty} U^1(t)=\mathcal{C}$ indicates that $U^1$ converges to not $U^1(0)$ but $\mathcal{C}$, because $\lim _{t \rightarrow \infty} {K_1(\chi(t))}/{K_2(\chi(t))}=1$ is obtained by $\lim _{t \rightarrow \infty} \chi(t)=\infty$ owing to inelastic relativistic-collisions.}\\\\
Next, we consider the cooling process by inelastic relativistic-collisions. Here, we consider the time evolution of $T^{0\alpha\beta}$ instead of that of $T^{\alpha\beta}$ to remove terms divided by $p^0$ in Eq. (7). Additionally, we assume that the distribution function is expressed by Maxwell-J$\ddot{\mbox{u}}$ttner function, namely, $f=f_{MJ}$.\\
The time evolution of $T^{0\alpha \beta}_E=\int_{-\infty}^\infty p^0 p^\alpha p^\beta f_{MJ} dp/p^0$ is written as
\begin{eqnarray}
d_t T^{0\alpha \beta}_E=\frac{A}{2}\int_{-\infty}^{\infty}\int_{-\infty}^{\infty} \left({p^\alpha}^\prime{p^\beta}^\prime+{p^\alpha_\ast}^\prime {p^\beta_\ast}^\prime-p^\alpha p^\beta-p^\alpha_\ast p^\beta_\ast \right) f_{MJ}\left(t,p\right) f_{MJ}\left(t,p_\ast\right) F \frac{dp_\ast}{p^0_\ast} \frac{d p}{p^0},
\end{eqnarray}
We introduce two vectors $P^\alpha$ and $Q^\alpha$, which are defined by
\begin{eqnarray}
&&P^\alpha=p^\alpha+p^\alpha_\ast,~~~~{P^{\alpha}}^\prime={p^\alpha}^\prime+{p^\alpha_\ast}^\prime, \nonumber \\
&&Q^\alpha=p^\alpha-p^\alpha_\ast,~~~~{Q^{\alpha}}^\prime={p^\alpha}^\prime-{p^\alpha_\ast}^\prime,
\end{eqnarray}
where we remind that $p^1+p^1_\ast={p^1}^\prime+{p^1_\ast}^\prime$ and $p^0+p^0_\ast \ge {p^0}^\prime+{p^0_\ast}^\prime$.\\
We obtain following relations from Eq. (11)
\begin{eqnarray}
&&P^\alpha Q_\alpha=0,~~~{P^\alpha}^\prime {Q_\alpha}^\prime=0, \\
&&P^2=P^\alpha P_\alpha=4+Q^\alpha Q_\alpha=4+Q^2.
\end{eqnarray}
From inverse transformation of Eq. (11), we obtain
\begin{eqnarray}
p^\alpha&=&\frac{1}{2}\left(P^\alpha+Q^\alpha\right),~~~p^\alpha_\ast=\frac{1}{2}\left(P^\alpha-Q^\alpha\right),\nonumber \\
{p^\alpha}^\prime&=&\frac{1}{2}\left({P^\alpha}^\prime+{Q^\alpha}^\prime\right),~~~{p^\alpha_\ast}^\prime=\frac{1}{2}\left({P^\alpha}^\prime-{Q^\alpha}^\prime\right),
\end{eqnarray}
Substituting Eq. (14) into Eq. (10), we obtain
\begin{eqnarray}
&&d_t T^{0\alpha \beta}_E= \nonumber \\
&&\frac{A}{8}\int_{-\infty}^{\infty}\int_{-\infty}^{\infty} \left[\left({P^\alpha}^\prime{P^\beta}^\prime-P^\alpha P^\beta \right)+\left( {Q^\alpha}^\prime{Q^\beta}^\prime-Q^\alpha Q^\beta \right)\right] f_{MJ}\left(t,p\right) f_{MJ}\left(t,p_\ast\right) g_{\o} dP dQ,
\end{eqnarray}
where we used the relation $F dp/p_0 dp_\ast/{p_0}_\ast=1/2 g_{\o} dPdQ$ \cite{Cercignani}.\\
The integration of ${A}/{8}\int_{-\infty}^{\infty}\int_{-\infty}^{\infty} \left({P^\alpha}^\prime{P^\beta}^\prime-P^\alpha P^\beta \right) f_{MJ}\left(t,p\right) f_{MJ}\left(t,p_\ast\right) g_{\o} dP dQ$ in the right hand side of Eq. (15) is markedly difficult, whereas ${P^1}^\prime{P^1}^\prime-P^1 P^1=0$.\\
Hereafter, we can neglect the integration ${A}/{8}\int_{-\infty}^{\infty}\int_{-\infty}^{\infty} \left({P^\alpha}^\prime{P^\beta}^\prime-P^\alpha P^\beta \right) f_{MJ}\left(t,p\right) f_{MJ}\left(t,p_\ast\right) g_{\o} dP dQ$ with $\alpha=0$ or $\beta=0$ in the right hand side of Eq. (15), because we will investigate the time evolution of $T^{011}$ in later discussion.\\
The center of mass system yields relations using Eqs. (2) and (3) 
\begin{eqnarray}
P^\alpha=\left(P^0,0\right),~~~Q^\alpha=\left(0,Q\right),~~~{Q^\alpha}^\prime=-\Lambda\left(0,Q\right).
\end{eqnarray}
From Eq. (16), we obtain
\begin{eqnarray}
{Q^\alpha}^\prime{Q^\beta}^\prime-Q^\alpha Q^\beta=-Q^2\left(1-\Lambda^2\right)
\left(
\begin{array}{cc}
0 & 0 \\
0 & 1 
\end{array}
\right)=-Q^2\left(1-\Lambda^2\right)\left(\frac{P^\alpha P^\beta}{P^2}-\eta^{\alpha \beta} \right),
\end{eqnarray}
where $\eta^{\alpha\beta}=\mbox{diag}(1,-1,-1,-1)$.\\
M\o ller's relative velocity in the center of mass system is \cite{Cercignani}
\begin{eqnarray}
g_{\o}=2 \frac{Q}{P^0}.
\end{eqnarray}
From Eqs. (17) and (18), we rewrite Eq. (15), when $\alpha=\beta=1$, as
\begin{eqnarray}
&& d_t T_E^{011}\nonumber \\
&=&\frac{A}{8}\int_{-\infty}^{\infty}\int_{-\infty}^{\infty} \left({Q^1}^\prime{Q^1}^\prime-Q^1 Q^1 \right) f_{MJ}\left(t,p\right) f_{MJ}\left(t,p_\ast\right) g_{\o} dP dQ \nonumber \\
&=&-\left(1-\Lambda^2\right)\frac{A}{4}\int_{-\infty}^{\infty}\int_{-\infty}^{\infty} \left(\frac{P^1 P^1}{P^2}-\eta^{11} \right) Q^3 f_{MJ}\left(t,p\right) f_{MJ}\left(t,p_\ast\right) \frac{dP}{P^0}dQ \nonumber \\
&=&-\left(1-\Lambda^2\right) n^2 \frac{A}{16 K_1(\chi)^2}\int_{-\infty}^{\infty} \left(\frac{{Z^\star}^{11}}{4+Q^2}-\eta^{11} Z^\star \right) Q^3 dQ\nonumber \\
&=&-\left(1-\Lambda^2\right) n^2 \frac{A}{4 K_1(\chi)^2} \int_{2}^{\infty} \left(K_2\left(\chi Q^\star\right) U^1 U^1 Q^\star-\eta^{11} \frac{K_1\left(\chi Q^\star\right)}{\chi}-\eta^{11} K_0 \left(\chi Q^\star\right) Q^\star \right) \left({Q^\star}^2-4\right) d Q^\star \nonumber \\
&=&-\left(1-\Lambda^2\right) n^2 \frac{A}{4 \chi K_1(\chi)^2} \int_{2\chi}^{\infty} \left(K_2\left(x\right) U^1 U^1 \frac{x}{\chi}-\eta^{11} \frac{K_1\left(x\right)}{\chi}-\eta^{11} K_0 \left(\chi Q^\star\right) \frac{x}{\chi} \right) \left(\frac{x^2}{\chi^2}-4\right) d x \nonumber \\
&=&-\left(1-\Lambda^2\right) n^2 \frac{A}{\chi^3 K_1(\chi)^2} \left[2\chi K_2\left(2\chi\right) U^1 U^1-\eta^{11} \left(2 \chi K_0(2\chi)+K_1(2\chi)\right)\right],
\end{eqnarray}
where $Z^\star$ and ${Z^\star}^{\alpha\beta}$ are defined in Eqs. (33) and (34), $Q^\star=\sqrt{Q^2+4}$ and $x=Q^\star/\chi$.\\
From Eqs. (8) and (19), we obtain the time evolution of $T^{011}_E$ as
\begin{eqnarray}
d_t T_E^{011}&=&-\left(1-\Lambda^2\right) n^2 \frac{A}{\chi^3 K_1(\chi){}^2} \left[2\chi K_2\left(2\chi\right) \left(U^1\right)^2 -\eta^{11} \left(2 \chi K_0(2\chi)+K_1(2\chi)\right)\right],\nonumber \\
&=& -\left(1-\Lambda^2\right) n^2 \frac{A}{\chi^3 K_1(\chi){}^2} \left[2\chi K_2\left(2\chi\right) \left(\mathcal{C}\frac{K_1(\chi)}{K_2(\chi)}\right)^2 -\eta^{11} \left(2 \chi K_0(2\chi)+K_1(2\chi)\right)\right].
\end{eqnarray}
We obtain following relation using Eqs. (32) and (8)
\begin{eqnarray}
d_t T^{011}_E&=&\frac{n}{\sqrt{1+\mathcal{C}^2 K_1(\chi)^2/K_2(\chi)^2}} \nonumber \\
&&\left[2 \chi ^2 \left(\mathcal{C}^2 \chi^4+8\mathcal{C}^2 \chi^2+3\chi^2+8 \right) K_1(\chi ){}^3 K_0(\chi ){}^2+2 \chi  \left(6\chi ^2-\mathcal{C}^2 \chi ^2 \left(\chi^2-8\right)+24\right) K_1(\chi ){}^4 K_0(\chi) \right. \nonumber \\
&&\left. -2 \left(\mathcal{C}^2 \chi^4 \left(\chi ^2+6\right)-4 \chi^2-16\right) K_1(\chi ){}^5+\chi ^3 \left(4 \mathcal{C}^2\chi ^2+\chi ^2-8\right) K_1(\chi ){}^2 K_0(\chi ){}^3 \right. \nonumber \\
&& \left. -\chi ^5 K_0(\chi){}^5-6 \chi ^4 K_1(\chi ) K_0(\chi ){}^4\right]\left(\chi ^4 K_1(\chi ){}^2 K_2(\chi){}^3\right)^{-1} \frac{d \chi^{-1}}{d t},
\end{eqnarray}
where we used $d_t \left(n U^0\right)=0$.\\
From Eqs. (20) and (21), we obtain
\begin{eqnarray}
d_t \chi^{-1}&=&-n A \left(1-\Lambda^2\right) \psi_1\left(\chi,\mathcal{C}\right) \chi^{-1}, \nonumber \\
\psi_1\left(\chi,\mathcal{C}\right) &=&\left(1+\mathcal{C}^2 K_1(\chi)^2/K_2(\chi)^2 \right)^{-\frac{1}{2}} \left[\chi^2 K_2(\chi) \left(2 \mathcal{C}^2 \chi K_2(2 \chi ) K_1(\chi ){}^2+(2 \chi  K_0(2 \chi )+K_1(2 \chi )) K_2(\chi ){}^2\right)\right] \nonumber \\
&& \left[2 \chi ^2 \left(\mathcal{C}^2 \chi^4+8 \mathcal{C}^2 \chi^2+3\chi^2+8\right) K_1(\chi ){}^3 K_0(\chi ){}^2+2 \chi \left(6\chi ^2-\chi ^4 \mathcal{C}^2+8 \chi ^2 \mathcal{C}^2+24\right) K_1(\chi){}^4 K_0(\chi) \right. \nonumber \\
&& \left. -2 \left(\mathcal{C}^2 \chi^6+6 \mathcal{C}^2 \chi ^4-4\chi ^2-16\right) K_1(\chi ){}^5+\chi ^3 \left(4 \mathcal{C}^2 \chi ^2+\chi ^2-8\right) K_1(\chi ){}^2 K_0(\chi ){}^3 \right. \nonumber \\
&& \left. -\chi ^5 K_0(\chi){}^5-6 \chi ^4 K_1(\chi ) K_0(\chi ){}^4\right]^{-1}.
\end{eqnarray}
From Eq. (22), we find that the cooling rate parameter $\psi_1\left(\chi,\mathcal{C}\right)$ depends on the frame via $u(0)$.\\
The left frame of Fig. 1 shows $\psi_1\left(\chi,\mathcal{C}\right)$ versus $\chi$ and $\mathcal{C}$. $\psi_1\left(\chi,\mathcal{C}\right)$ is symmetric at both sides of $\mathcal{C}=0$. $\psi_1\left(\chi,\mathcal{C}\right)$ approximates to 0, when $\chi$ approximates to $\infty$, whereas $\psi_1\left(\chi,\mathcal{C}\right)$ approximates to $1/8$, when $\chi$ approximates to 0. Additionally, $\psi_1\left(\chi,\mathcal{C}\right)$ has a peak, which moves toward to $\chi=0$, as $|\mathcal{C}|$ increases.\\
Provided that $\mathcal{C}=0$ ($u(0)=0$), we obtain $\psi_1\left(\chi,0\right)$ from Eq. (22) as
\begin{eqnarray}
\psi_1\left(\chi,0\right)=\frac{2 \chi  K_0\left(2 \chi\right)+K_1\left(2 \chi\right)}{\chi \left(\chi ^2+4\right) K_1\left(\chi\right){}^2-\chi ^3 K_0\left(\chi\right){}^2}.
\end{eqnarray}
The right frame of Fig. 1 shows $\psi_1\left(\chi,0\right)$ versus $\chi$. $\psi_1\left(\chi,0\right)$ has its peak value $0.295$ at $\chi \simeq 4.14$, whereas we obtain $\lim_{\chi \rightarrow 0} \psi_1(\chi,0)=1/8$ and $\lim_{\chi \rightarrow \infty} \psi_1(\chi,0)=2/\sqrt{\pi \chi}$, which coincides with the cooling rate parameter, which is calculated using 1D inelastic nonrelativistic-Boltzmann equation.
\begin{center}
\includegraphics[width=0.9\textwidth]{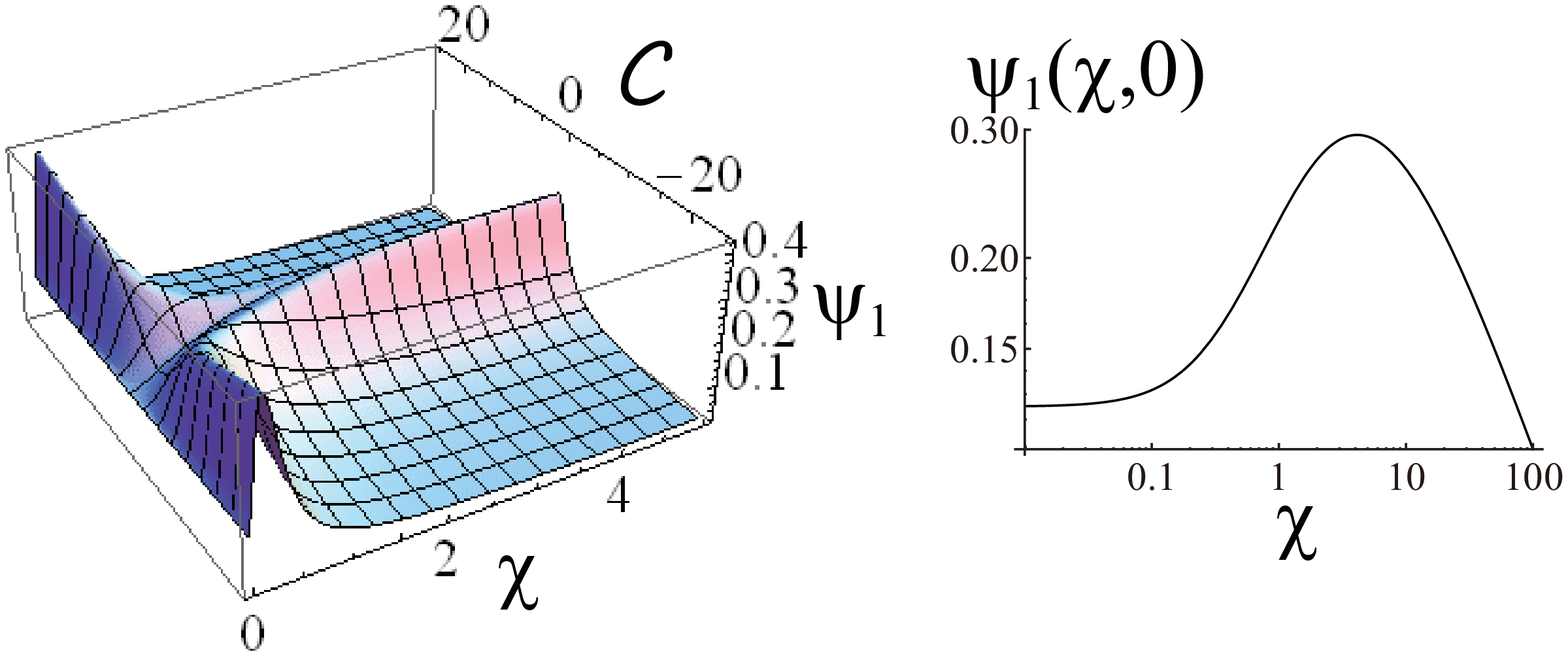}\\
\footnotesize{FIG. 1: $\psi_1\left(\chi,\mathcal{C}\right)$ versus $\chi$ and $\mathcal{C}$ (left frame), and $\psi_1\left(\chi,0\right)$ versus $\chi$ (right frame).}
\end{center}
\section{Numerical analysis of cooling process}
We investigate the characteristics of the cooling process, which is derived from the inelastic relativistic-collision between two quarks, numerically.\\
Firstly, we set $\Lambda=0$ and $A=1$ in Eq. (1). From Eq. (22), we know that the cooling rate depends on $\chi$ and $\mathcal{C}$, when $f=f_{MJ}$. As initial data, we consider two tests. One is Test 1, in which $f$ is uniformly populated in the range of $0.99 \le \left|v\right|<1$ at $t=0$. The other is Test 2, in which $f$ is uniformly populated in the range of $0.99\le v <1$ and $-1< v \le -0.9$ at $t=0$. As a result, initial distribution functions in Tests 1 and 2 are markedly nonequilibrium. Meanwhile, DSMC sampling of the initially equilibrium state $f=f_{MJ}$ is markedly difficult, when $u \neq 0$ and 1D elastic relativistic-collisions occur. Thus, we use the initially nonequilibrium state.\\
Figure 2 shows time evolutions of $\chi$, which are obtained in Tests 1 and 2. $\chi$ in Test 1 is smaller than $\chi$ in Test 2 at $t=0$, whereas $\chi$ in Tests 1 and 2 increase with similar inclinations in the range of $0 \le t \le 6$. The inclination of $\chi$ in Test 2 decreases at $t \simeq 6$. Provided that $f \sim f_{MJ}$ at $6 \le t$, $\psi_1\left(\chi,\mathcal{C}\right)$ in Test 2 is smaller than $\psi_1\left(\chi,\mathcal{C}\right)$ in Test 1. The cooling rate parameter $\psi_1\left(\chi,\mathcal{C}\right)$ increases, as $|\mathcal{C}|$ decreases, when $1<\chi$, as shown in Fig. 1. Provided that $|\mathcal{C}|$ in Test 1 is smaller than $|\mathcal{C}|$ in Test 2, $\psi_1\left(\chi,\mathcal{C}\right)$ in Test 1 is larger than $\psi_1\left(\chi,\mathcal{C}\right)$ in Test 2, when $f=f_{MJ}$ and $1<\chi$. In later discussion, we, however, find that $\mathcal{C}=U^1K_2\left(\chi\right)/K_1\left(\chi\right)$ is not the temporal constant owing to $f \neq f_{MJ}$, because $T^{01}$ $\left(\neq T_E^{01}\right)$ includes effects of nonequilibrium moments such as $\Pi$, $q^\alpha$, and $\Pi^{\alpha\beta}$ \cite{Eckart}.
\begin{center}
\includegraphics[width=0.5\textwidth]{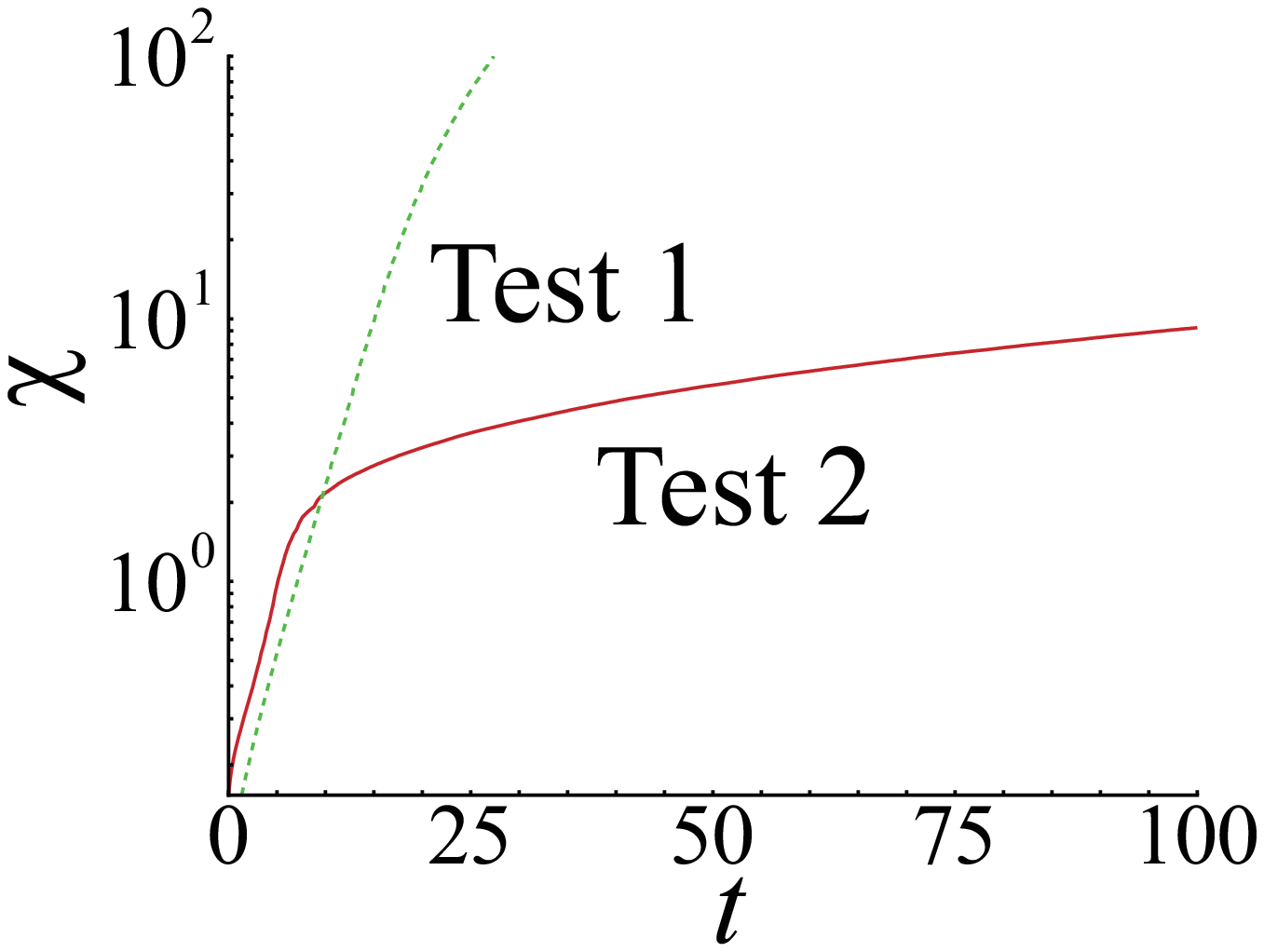}\\
\footnotesize{FIG. 2: time evolutions of $\chi$ in Tests 1 and 2.}
\end{center}
Figure 3 shows $f$ and $f_{MJ}$ versus $v$ at $t=17.7$ in Test 1 (left frame) and $f$ and $f_{MJ}$ versus $v$ at $t=177$ in Test 2 (right frame). The left frame of Fig. 3 shows that $f$ has higher tails than $f_{MJ}$ at $0.575 \le |v|$. Such higher tails are obtained under the homogeneous cooling state of the granular gas \cite{Ben-Naim}. Meanwhile, $f$ has a higher tail at $0.99 \le v <1$ and lower tail at $v \le 0.965$. Therefore, the overpopulation of $f$ versus $f_{MJ}$ at the high velocity tail in both sides of $v=u$, which is always obtained by the inelastic nonrelativistic-collisions for any value of $u$, is not obtained, when $u$ is similar to the speed of light, as shown in the right frame of Fig. 3. The initial marked differences between $f$ and $f_{MJ}$ are reduced by inelastic relativistic-collisions in both Tests 1 and 2.
\begin{center}
\includegraphics[width=0.8\textwidth]{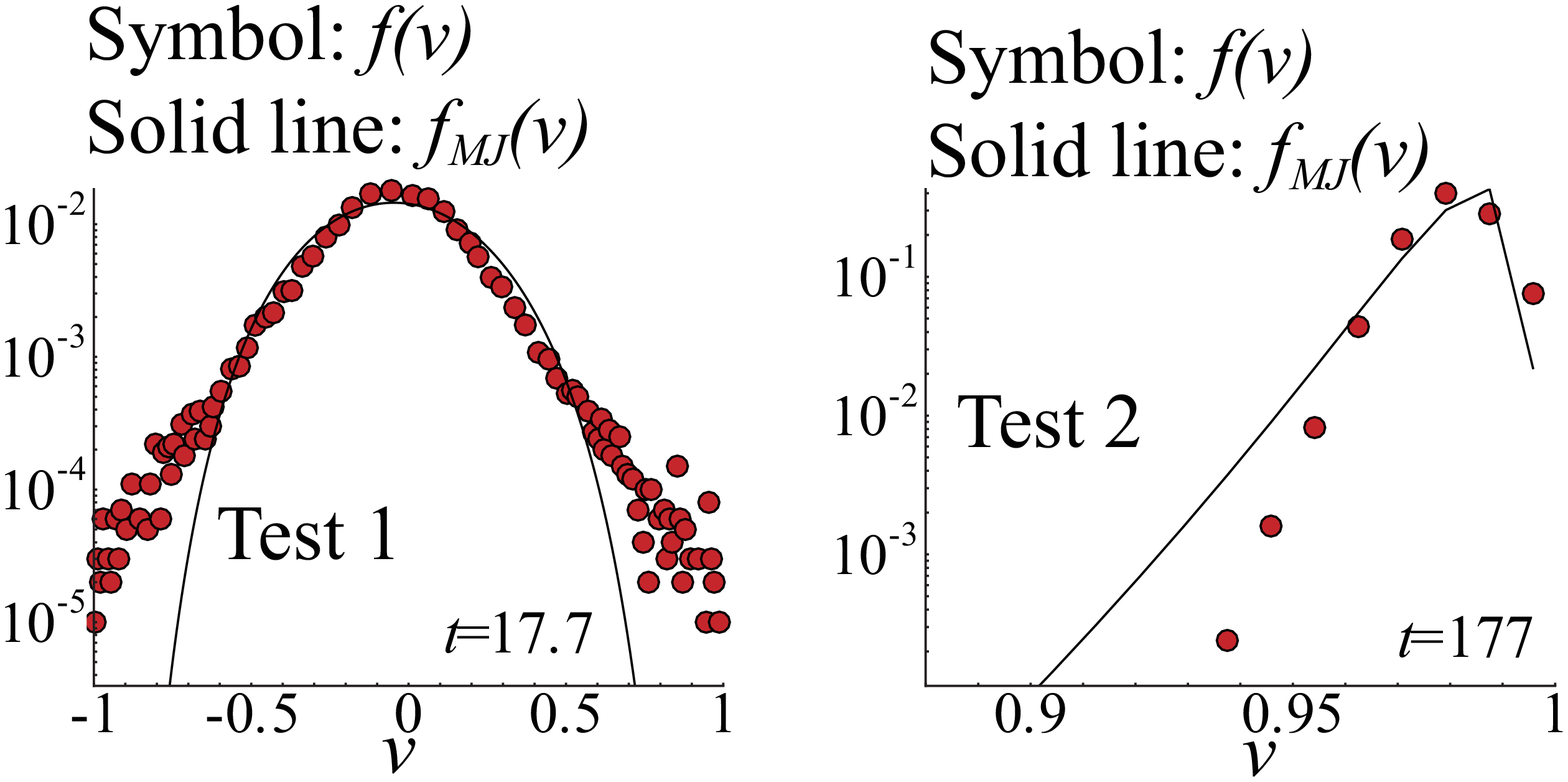}\\
\footnotesize{FIG. 3: $f(v)$ and $f_{MJ}(v)$ versus $v$ at $t=17.7$ in Test 1 (left frame). $f(m)$ and $f_{MJ}(v)$ versus $v$ at $t=177$ in Test 2 (right frame).}
\end{center}
Next, we investigate the cooling process of $\theta$, which is analytically obtained in Eq. (22). Meanwhile, the calculation of the general solution of $\chi$ in Eq. (22) is markedly difficult. Then, we calculate the limiting solution of $\chi$ in Eq. (22). Provided that $\mathcal{C}=0$ ($u_{t=0}=0$) in Eq. (22), we obtain solutions of $\chi$ under two limiting cases, namely, $\chi \rightarrow 0$ and $\chi \rightarrow \infty$ from Eq. (23) as
\begin{eqnarray}
\chi(t)&=&\chi(0) \exp\left(\frac{A}{8} t\right)~~~~(\chi \rightarrow 0). \\
&=&\left(A\frac{t}{\sqrt{\pi}}+\sqrt{\chi(0)}\right)^2~~~~(\chi \rightarrow \infty).
\end{eqnarray}
Equations. (24) and (25) are plotted together with $\chi$ in Test 1 in Fig. 4, where $A=1$ is used in Eqs. (24) and (25). $\chi_{t=0}$ in Eq. (24) is defined by $\chi_{t=0}$ in Test 1 and $\chi_{t=0}$ in Eq. (25) is defined by $\chi_{t=15}$ in Test 1. Eq. (24) gives a good agreement with $\chi$ in Test 1 in the range of $0 \le t \le 0.5$, whereas Eq. (25) gives a good agreement with $\chi$ in Test 1 in the range of $15 \le t$. Such good agreements are obtained, because initial nonequilibrium of $f$ is reduced by the inelastic relativistic-collisions, as shown in the left frame of Fig. 6.\\
Next, we investigate the time evolution of the flow velocity $u$ in Tests 1 and 2. Fig. 5 shows time evolutions of $u$ in Tests 1 and 2. $u$ in Test 1 temporally decreases, whereas $u$ in Test 2 temporally increases, as shown in the left frame of Fig. 5.
\begin{center}
\includegraphics[width=0.5\textwidth]{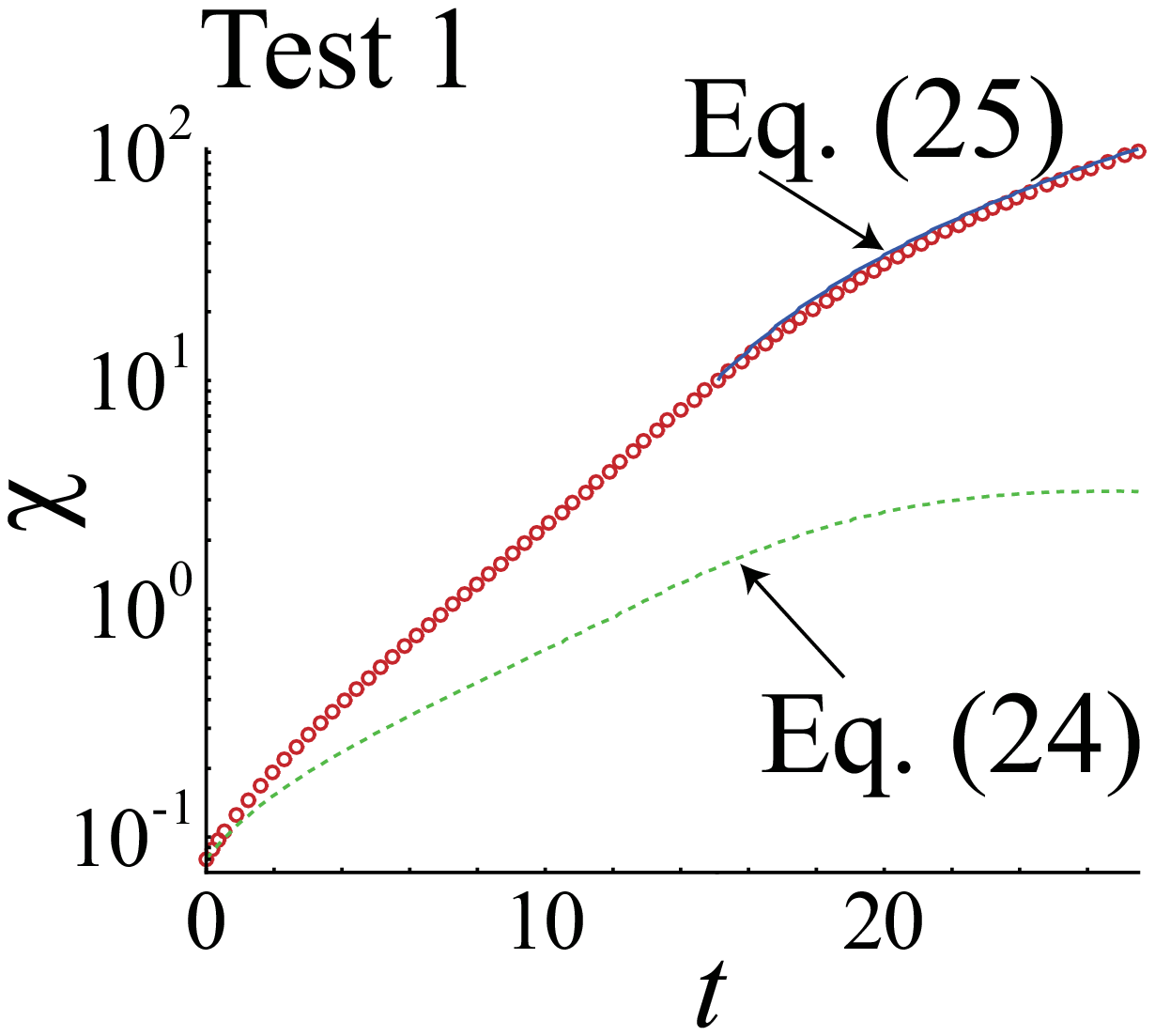}\\
\footnotesize{FIG. 4: Plots of Eqs. (24) and (25) together with time evolution of $\chi$ in Test 1 .}
\end{center}
As described in Remark 2.1, $|u|$ increases, as $\chi$ increases. Here, we must remind that $u_{t=0}=-\epsilon$ ($0<\epsilon \ll 1$) in Test 1, which is caused by the thermal fluctuation at $t=0$, yields the time evolution of $u$ in the range of $u_{t=0}<0$. Therefore, the signature of $u$ is determined by the signature of $u_{t=0}$. $u$ in Test 2 approximates to 0.97, as $t$ increases. Such a increase of $u$ via the cooling process of $\theta$ indicates that $|v|$ of quarks are accelerated by inelastic relativistic-collisions between two quarks, unless the initial flow velocity $u_{t=0}$ is zero, namely, $u_{t=0}=0$ and $f=f_{MJ}$ for $0 \le t$. In particular, the initial signature of $u_{t=0}$, namely, $0<u_{t=0}$ or $u_{t=0}<0$ coincides with the final signature of $\lim_{t \rightarrow \infty} u$. Of course, such a acceleration of $|u|$ is not obtained by inelastic nonrelativistic-collisions, because $u$ is always conserved at $0 \le t$. 
\begin{center}
\includegraphics[width=0.5\textwidth]{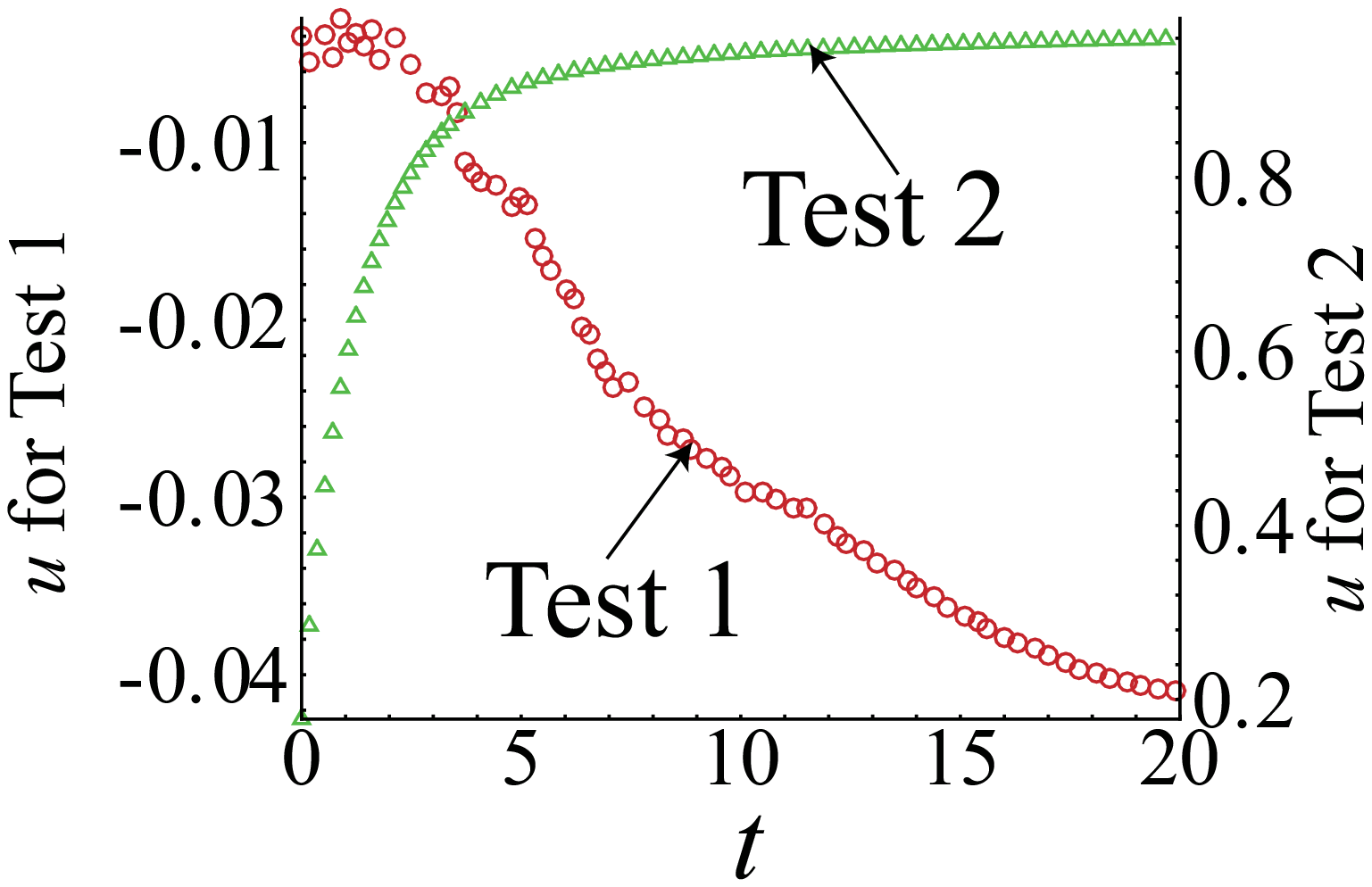}\\
\footnotesize{FIG. 5: Time evolutions of $u$ in Tests 1 and 2.}
\end{center}
Finally, we must mention to effects of nonequilibrium of $f$, briefly. Figure 6 shows time evolutions of $\Pi$, $q^1$, $\Pi^{\left<11\right>}$ and $U^1K_1(\chi)/K_2(\chi)$, which must be a temporally constant, namely, $\mathcal{C}$, when $f=f_{MJ}$, in Test 1 (left frame) and Test 2 (right frame). As shown in the left frame of Fig. 6, $U^1K_1(\chi)/K_2(\chi)$ increases in the range of $0 \le t \le 12$ owing to nonequilibrium effects and approximates to the constant value in the range of $12 <t$. Additionally, the left frame of Fig. 6 indicates that $\Pi$, $q^1$ and $\Pi^{\left<11\right>}$ are significant in the range of $0 \le t \le 12$ and temporally damped to zero in the range of $12<t$ with similar damping rates. As shown in the right frame of Fig. 6, $U^1K_1(\chi)/K_2(\chi)$ increases in the range of $0 \le t \le 2.5$ owing to nonequilibrium effects, decreases in the range of $2.5 < t \le 8.8$, and approximates to the constant value in the range of $8.8 <t$. Additionally, $\Pi$ and $q^1$ are temporally damped to 0 in the range of $6 \le t$ with similar damping rates, whereas $\Pi^{\left<11\right>}$ is temporally damped with a slower damping rate than damping rates of $\Pi$ and $q^1$.\\
Provided that we can assume $f \sim f_{MJ}$ at $6<t$ in Tests 1 and 2, $|U^1K_1(\chi)/K_2(\chi)|_{\mbox{\tiny{Test 1}}} \ll |U^1K_1(\chi)/K_2(\chi)|_{\mbox{\tiny{Test 2}}}$, which is obtained from the left and right frames of Fig. 6, indicates that $|\mathcal{C}|_{\mbox{\tiny{Test 1}}} \ll |\mathcal{C}|_{\mbox{\tiny{Test 2}}}$ might prove $(\psi_1)_{\mbox{\tiny{Test 2}}}<(\psi_1)_{\mbox{\tiny{Test 1}}}$ at $6<t$ in Fig. 2 by setting $t=6$ to $t=0$.
\begin{center}
\includegraphics[width=1.0\textwidth]{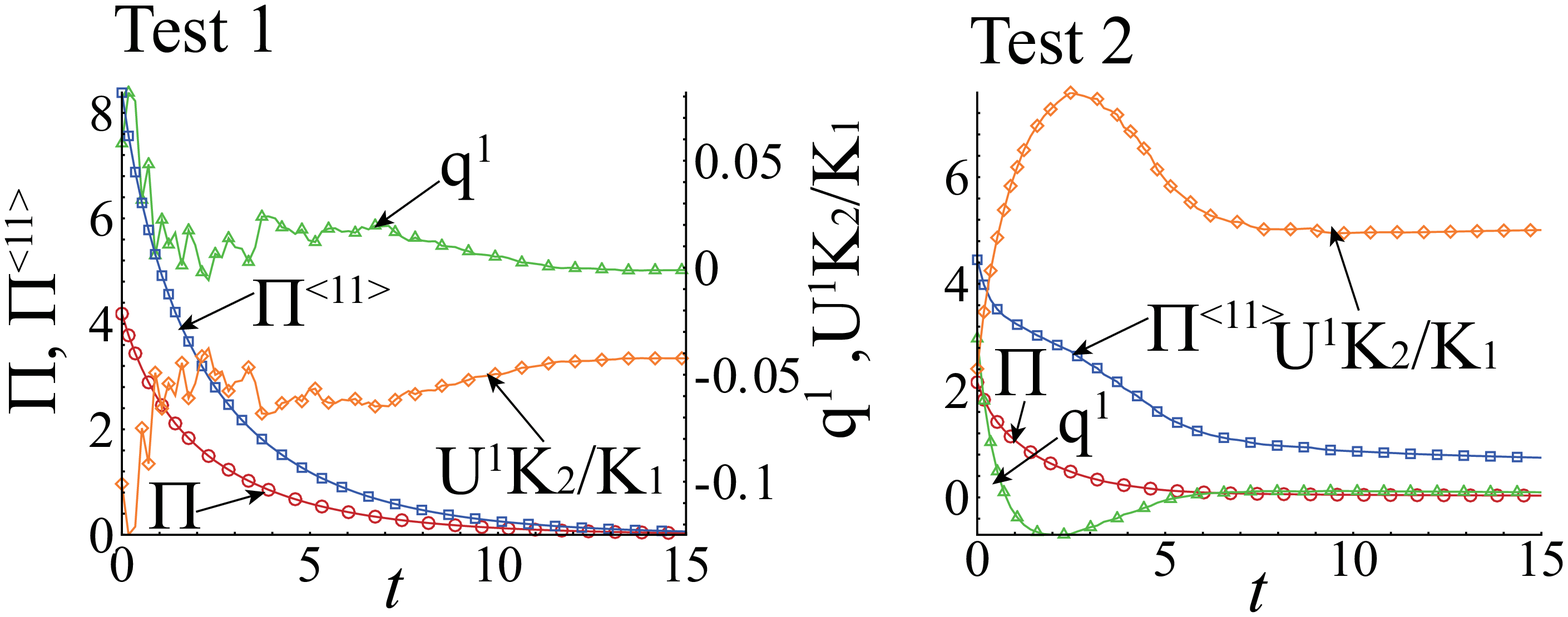}\\
\footnotesize{FIG. 6: time evolutions of $\Pi$, $q^1$, $\Pi^{11}$, and $U^1K_1(\chi)/K_2(\chi)$ in Test 1 (left frame) and Test 2 (right frame).}
\end{center}
\section{Concluding Remarks}
In this paper, we investigated time evolutions of the temperature and flow velocity, when the binary collision between two quarks is described by 1D inelastic relativistic-Boltzmann equation. The absolute value of the flow velocity increases by the decrease in the temperature owing to the inelastic relativistic-collisions, whereas the signature of the flow velocity at $0<t$ coincides with that of the initial flow velocity. The cooling rate of the temperature depends on both $\chi$ ($\theta$) and $\mathcal{C}$ in Eq. (9), which is the function of the initial temperature and flow velocity, under the equilibrium state. To confirm the analytical results, we solved 1D inelastic relativistic-Boltzmann equation under the spatially homogeneous state. Numerical results confirm that the time evolution of the temperature, which is analytically obtained in Eqs. (24) or (25), gives good agreements with numerical results of the time evolution of the temperature. Of course, the cooling rate depends on nonequilibrium moments such as the dynamics pressure, pressure deviator, heat flux and so on, which are nonzero values during calculations, as shown in Fig. 6. Additionally, the numerical results surely confirm that the flow velocity increases by the decrease in the temperature owing to the inelastic relativistic-collisions. Finally, the distribution function does not show the overpopulation in the high velocity tail at both sides of the flow velocity (i.e., $|u| \ll |v|$), when the flow velocity approximates to the speed of light. Such a tendency of the distribution function is not obtained under the nonrelativistic limit, because the distribution function always shows the overpopulation in the high velocity tail at both sides of the flow velocity (i.e., $|u| \ll |v|$) under the spatially homogeneous state.
\begin{appendix}
\section{Definitions of equilibrium moments}
In this appendix, some equilibrium moments are defined.\\
At first, the zeroth order moment is defined as
\begin{eqnarray}
Z=\int_{-\infty}^{\infty} \exp(-\chi p^\alpha U_\alpha) \frac{dp}{p^0}.
\end{eqnarray}
In Lorentz rest frame, we obtain
\begin{eqnarray}
Z=\int_{-\infty}^\infty \exp\left(-\chi p^0 U_0\right) \frac{dp}{p^0}=2 K_0(\chi).
\end{eqnarray}
$Z^{\alpha\beta\gamma\delta...}=\int_{-\infty}^\infty p^\alpha p^\beta p^\gamma p^\delta ... \exp(-\chi p^\alpha U_\alpha) \frac{dp}{p^0}$ is obtained by the successive differentiation of $Z$ using $-\chi U_\alpha$ as
\begin{eqnarray}
&&Z^\alpha=2 K_1(\chi)U^\alpha,\\
&&Z^{\alpha\beta}=2 K_2(\chi)U^\alpha U^\beta-2\eta^{\alpha\beta} \frac{K_1(\chi)}{\chi},\\
&&Z^{\alpha\beta\gamma}=2 K_3(\chi)U^\alpha U^\beta U^\gamma-2\left(\eta^{\alpha\beta}U^\gamma+\eta^{\alpha\gamma}U^\beta+\eta^{\beta\gamma}U^\alpha \right)\frac{K_2(\chi)}{\chi},\\
&&Z^{\alpha\beta\gamma\delta}=2 K_4(\chi)U^{\alpha} U^\beta U^\gamma U^\delta \nonumber \\
&&-\frac{2 K_3(\chi)}{\chi}\left(\eta^{\alpha\beta} U^\gamma U^\delta+\eta^{\alpha\gamma} U^\beta U^\delta+\eta^{\beta\gamma} U^\alpha U^\delta+\eta^{\alpha\delta} U^\gamma U^\beta+\eta^{\delta\gamma} U^\beta U^\alpha+\eta^{\delta\beta} U^\alpha U^\gamma\right) \nonumber \\
&&+\frac{2 K_2(\chi)}{\chi^2}(\eta^{\alpha\beta}\eta^{\gamma\delta}+\eta^{\alpha\gamma}\eta^{\beta\delta}+\eta^{\alpha\delta}\eta^{\beta\gamma}),
\end{eqnarray}
Similarly, ${Z^\star}^{\alpha\beta\gamma\delta...}=\int_{-\infty}^\infty P^\alpha P^\beta P^\gamma P^\delta ... \exp(-\chi P^\alpha U_\alpha) \frac{dP}{P^0}$ is obtained by the successive differentiation of $Z^\star$ using $-\chi U_\alpha$ as
\begin{eqnarray}
&&Z^\star=2 K_0\left(Q^\star \chi \right),\\
&&{Z^\star}^\alpha=2 Q^\star K_1(Q^\star \chi)U^\alpha,\\
&&{Z^\star}^{\alpha\beta}=2 {Q^\star}^2 K_2(Q^\star \chi)U^\alpha U^\beta-2 Q^\star \eta^{\alpha\beta} \frac{K_1(Q^\star \chi)}{\chi},
\end{eqnarray}
\end{appendix}

\end{document}